\documentclass[preprint,amsmath,superscriptaddress,showpacs,showkeys,amssymb,aps,prb]{revtex4-1}
\usepackage{graphicx,epsfig,multirow,dcolumn,bm}

\begin{document}
\title{Mechanism of magnetostructural transformation in multifunctional Mn$_3$GaC}

\author{E. T. Dias}
\author{K. R. Priolkar}
\affiliation{Department of Physics, Goa University, Taleigao Plateau, Goa 403206 India}
\email{krp@unigoa.ac.in}
\author{Rajeev Ranjan}
\affiliation{Department of Materials Engineering, Indian Institute of Science, Bangalore 560012, India}
\author{A. K. Nigam}
\affiliation{Tata Institute of Fundamental Research, Dr. Homi Bhabha Road, Colaba, Mumbai 400005, India}
\author{S. Emura}
\affiliation{Institute of Scientific and Industrial Research, Osaka University, Osaka, Japan}

\begin{abstract}
Mn$_3$GaC undergoes a ferromagnetic to antiferromagnetic, volume discontinuous cubic-cubic phase transition as a function of temperature, pressure and magnetic field. Through a series of temperature dependent x-ray absorption fine structure spectroscopy experiments at the Mn K and Ga K edge, it is shown that the first order magnetic transformation in Mn$_3$GaC is entirely due to distortions in Mn sub-lattice and with a very little role for Mn-C interactions. The distortion in Mn sub-lattice results in long and short Mn-Mn bonds with the longer Mn-Mn bonds favoring ferromagnetic interactions and the shorter Mn-Mn bonds favoring antiferromagnetic interactions. At the first order transition, the shorter Mn-Mn bonds exhibit an abrupt decrease in their length resulting in an antiferromagnetic ground state and a strained lattice.
\end{abstract}
\date{\today}

\maketitle

\section{Introduction}
Geometric frustration in lattices with triangular geometries and antiferromagnetic interactions refers to the inability of the three spins to simultaneously align antiparallel to each other. In systems such as these, the material orders by undergoing a structural phase transition to an unusual magnetic state or simply enters a glassy like state with decreasing temperatures \cite{Henley198962,Chubukov199373,Gaulin199485,Bergman20073}. In recent times, the possibility of fine tuning and controlling functional properties of geometrically frustrated Mn based antiperovskites has attracted considerable attention in terms of both fundamental research and potential applications \cite{Li200572,Tong200673,Tong200674,Tong2006138,Wang200995,Wang2009105,Wang201097,Lin2010108,Takenaka200587,Takenaka2008992,Chi2001120,Lin201199,PhysRevB.95.144418}. Among the others, technologically useful phenomena that arise in these compounds at temperatures associated with their respective first order magnetostructural transitions include large magnetocaloric effects (MCE)\cite{Tohei200394,Lewis200393,Aczel201490} and a giant magnetoresistance (GMR) of more than 10\% in pulsed magnetic field \cite{Kamishima200063,Li200572,Zhang2014115} in Mn$_{3}$GaC, the invar effect or even giant negative thermal expansion in Mn$_{3}$NiN \cite{Wu2013114} and near zero temperature coefficient of resistance in nitride (Mn$_{3}$AN, A = Ga, Zn, Cu) compounds\cite{Chi2001120,Sun201062}, magnetostriction in Mn$_{3}$CuN \cite{Shibayama2011109} and piezomagnetic effects in Mn$_{3}$GaN \cite{Lukashev200878}.

Extensive studies on multifunctional antiperovskite compound Mn$ _{3} $GaC have been carried out with the hope of understanding the highly correlated behavior of magnetism and structure exhibited by the material \cite{Kanomata199867,Meenakshi2006140,Dias2014363}. Crystallographic studies report that the intermetallic compound has a cubic structure belonging to space group Pm$\bar 3$m ($a$ = 3.896 \AA~ at 291 K) analogous to a perovskite but with the positions of cations and anions interchanged in the general formula ABX$ _{3} $. Within the cubic unit cell formed by Ga atoms positioned at the corners of the cube, Mn atoms occupy the face centered positions forming a corner sharing Mn$ _{6} $C octahedra with the C atom positioned at the center of the cube \cite{Bouchaud196637}.

Magnetically the compound is known to exhibit three phases; when cooled below its Curie temperature (T$ _{C} $ $\sim$242 K), it transforms from a room temperature paramagnetic (PM) to ferromagnetic (FM) state and upon further cooling, the FM component abruptly disappears and the compound transforms to an antiferromagnetic (AFM) ground state via a first order transition at about 178 K which is characterized as the N\'eel temperature T$ _{N} $. The magnetic crossover from the FM to AFM state is accompanied by a volume discontinuous cubic to cubic structural transition at T$ _{N} $. The AFM state has a higher unit cell volume ($\sim$ 0.5\%) than the FM state \cite{Dias2014363,Bouchaud196637,Fruchart19708,Kim2001119}. Alternation of ferromagnetically aligned Mn spins in consecutive (111) planes along the [111] direction results in the high volume AFM structure described by propagation vector $k = [\frac{1}{2}, \frac{1}{2}, \frac{1}{2}]$\cite{Fruchart19708}.

Goodenough-Anderson-Kanomari rules \cite{Goodenough1963} have been invoked to explain magnetic interactions present in Mn$_3$GaC. According to these rules, spin correlations between t$ _{2g} $ electrons of Mn atoms along the edges of the Mn$ _{6} $C octahedra shown in the inset of Figure \ref{fig:all}a provide 90$^\circ$ Mn-Mn ($J1$) FM interactions while the stronger Mn-C-Mn ($J2$) AFM interactions arise from the pairing of half-filled $p$ or $sp^3$ hybridized orbitals of C atoms with those of near neighbor Mn e$_{g}$ electrons \cite{Hua201024}. It is this competition between the two exchange interactions that is believed to be responsible for magnetic ground state in Mn$_3$GaC \cite{Guillot1964258,Fruchart197844}. However, C K edge x-ray absorption experiments have failed to show any changes in the Mn $3d$ and C $2p$ hybridization at the first order transition from FM to AFM state \cite{Lewis200618}. Furthermore, though experiments have shown the ground state of Mn$ _{3} $GaC to be cubic antiferromagnet, band structure studies reveal that the FM phase has a slightly lower (by $\sim$ 70 meV) formation energy as compared to the AFM phase \cite{Shim200266}. This negligible difference in energy strongly reflected as a competition between two phases that have nearly equal density states at the Fermi level (E$ _{F} $) makes the compound magnetically unstable \cite{Kanomata199867,Shim200266,Kaneko198756,Cakir2014115}.

High pressure and magnetic field dependent measurements bring out the most salient feature of the antiperovskite compound. For instance magnetic fields $ \geq5$ T can suppress/aid spin flip transitions and induce ferromagnetism even in the AFM state \cite{Kanomata199867,Kim2001119,Dias2014363,Cakir2014115}. The field induced AFM to FM transition in magnetic field is also accompanied by a reverse volume change \cite{Cakir2014115}. Identical results like enhanced Mn moment, stabilized FM phase and subsequent increase in T$ _{C} $ have been reported from high pressure studies \cite{Meenakshi2006140,Kaneko198756}. A direct reflection of this temperature and magnetic field induced first order transition are the technologically significant properties such as giant magnetoresistance and large and reversible inverse magnetocaloric effect with a table like field dependence \cite{Dias2014363,Kim2001119,Cakir2012100,Dias2015117}.

The temperature and field induced metamagnetic transitions in Mn$_3$GaC have been discussed in early literature by considering presence of trigonal strain or distortions that lower the symmetry from cubic (m$\bar3$m) to rhombohedral ($\bar3$m) \cite{Michalski197887,Cofta1982114}. However, no structural change has been reported in Mn$_3$GaC so far. Therefore there is a need to understand the magneto-structural transition in Mn$ _{3} $GaC afresh. In particular, a proper mechanism explaining the temperature and magnetic field and pressure induced AFM to FM transition in Mn$_3$GaC is needed.

In this work we systematically study the changes in the local structures of Mn and Ga as a function of temperature by x-ray absorption fine structure (XAFS) spectroscopy and correlate them with magnetic properties of Mn$_3$GaC. Our results clearly show a presence of distortion in the Mn sublattice. The Mn atoms are displaced from their face centered positions resulting in long and short Mn-Mn bond distances. While the longer Mn-Mn distances support ferromagnetic order, the shorter Mn-Mn distance supports AFM ordering. The magnetic ground state is then decided by a competition between these two Mn-Mn distances. The AFM ground state results due to a sharp decrease in the shorter Mn-Mn distance at the first order magnetic transition temperature. The lattice instability caused by this decrease in Mn-Mn distance is arrested by an increase in lattice volume. A compressive strain in the form of an applied magnetic field or hydrostatic pressure results in restoration of lattice volume, causing the shorter Mn-Mn distances to relax and reinduce the ferromagnetic order.

\section{Experimental section}
To synthesize polycrystalline Mn$ _{3} $GaC, stoichiometric proportions of elemental Mn and C powders were carefully mixed with ingots of Ga, pelletized and sealed in an evacuated quartz tube that was heated at 1073 K for five days. On cooling to room temperature the compound formed was pulverized and mixed with additional carbon powder (about 20\%) before being pressed into a pellet that was annealed using a similar procedure \cite{Dias2014363,Lewis200393}. Temperature dependent x-ray diffraction patterns were recorded using CuK$ \alpha $ radiation while magnetization measurements in the 5 K to 300 K range in fields upto 7 T were carried out using a Quantum Design SQUID magnetometer and Quantum Design Vibrating sample magnetometer. Absorbers used for EXAFS measurements in transmission mode were made up of multiple strips of scotch tape uniformly coated with powdered material that gave an absorption jump $\mu\leq$1. Ionization chambers filled with appropriate gases were used to simultaneously record the incident (I$  _{0}$) and transmitted (I$ _{t} $) photon intensities at the Ga (10367 eV) and Mn (6539 eV) K edges (using beamlines NW10A and BL-9C at Photon Factory, Japan and P-65 at PETRA-III synchrotron source, DESY, Hamburg, Germany) in Mn$ _{3} $GaC. Extraction of structural parameters was carried out using standard procedures such as pre-processing of data, theoretical modeling with basic crystallographic information and fitting of the $R$-space experimental data to theoretical EXAFS functions using various applets of the Demeter program \cite{Ravel200512}.

\section{Results and Discussion}
\begin{figure}[h]
\begin{center}
\includegraphics[width=\columnwidth]{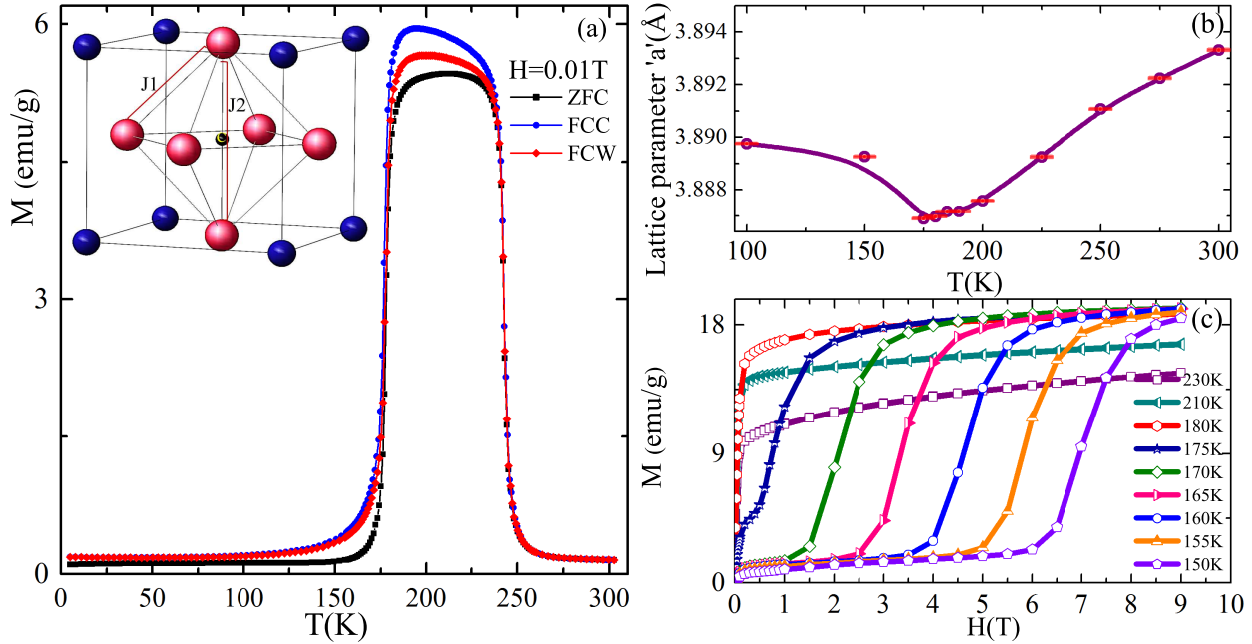}
\caption{Structural and Magnetic properties. (a) Magnetization as a function of temperature recorded in an applied field of 100 Oe during the ZFC (black squares), FCC (blue circles) and FCW (red diamonds) cycles. The inset shows a schematic of the unit cell and the two magnetic interactions denoted by J1 and J2 believed to be responsible for magnetic transitions in Mn$_3$GaC. (b) Variation of lattice parameter $`a'$ as a function of temperature obtained from x-ray diffraction measurements. (c) Isothermal magnetization recorded at different temperatures as a function of applied magnetic field. These properties are consistent with results reported earlier \cite{Bouchaud196637,Dias2014363,Kanomata199867}.}
\label{fig:all}
\end{center}
\end{figure}

Structural and magnetic properties of compound Mn$ _{3} $GaC prepared using the solid state reaction technique are consistent with earlier reports \cite{Dias2014363,Kanomata199867}. Temperature dependent magnetization recorded in the temperature range between 5 K to 300 K during zero field cooled (ZFC) and field cooled cooling (FCC) and subsequent warming (FCW) cycles is presented in Figure \ref{fig:all}(a). Transitions from the PM to FM state at 242 K via a second order transition and from FM to AFM state at 170 K via a first order transition are clearly visible in the magnetization data. Lattice parameter $a$ obtained from temperature dependent x-ray diffraction (XRD) patterns is plotted in Figure \ref{fig:all}(b). XRD patterns at some representative temperatures are depicted in supplementary text \cite{supp}. The lattice parameter shows a continuous monotonic decrease from 300 K to about 170 K below which there is an abrupt increase signalling expansion in unit cell volume at temperature coinciding with the first order transformation to the AFM ground state (T$ _{N} $ = 170 K). No distortions from cubic symmetry are noticed in the low temperature XRD data. Further, isothermal magnetization curves in Figure \ref{fig:all}(c) recorded as a function of applied field (M(H)) validate the field induced nature of the first order AFM-FM transition. At temperatures just below T$ _{C} $ and above T$ _{N} $ (viz. T = 180 K) the behavior of the magnetization curves is typical for the FM state. The nature of M(H) curves is non saturating does not saturate even in applied magnetic fields of 9 T. This non saturating nature of magnetization with increasing magnetic field indicates presence of competing antiferromagnetic interactions. The M(H) curves recorded when the compound is cooled below T$_N$,  (T $ \leq $ 170 K) perfectly highlights the field induced AFM to FM transition in Mn$ _{3} $GaC. In the AFM state, initially the magnetization has a low value before rapidly rising to the saturated FM state at higher applied fields. The field required to induce this first order transition continuously increases with decreasing temperatures. At 150 K, an applied field of 6 T is required to induce the FM state which increases to 20 T at 4 K \cite{Kanomata199867}.

\begin{figure}
\begin{center}
\includegraphics[width=\columnwidth]{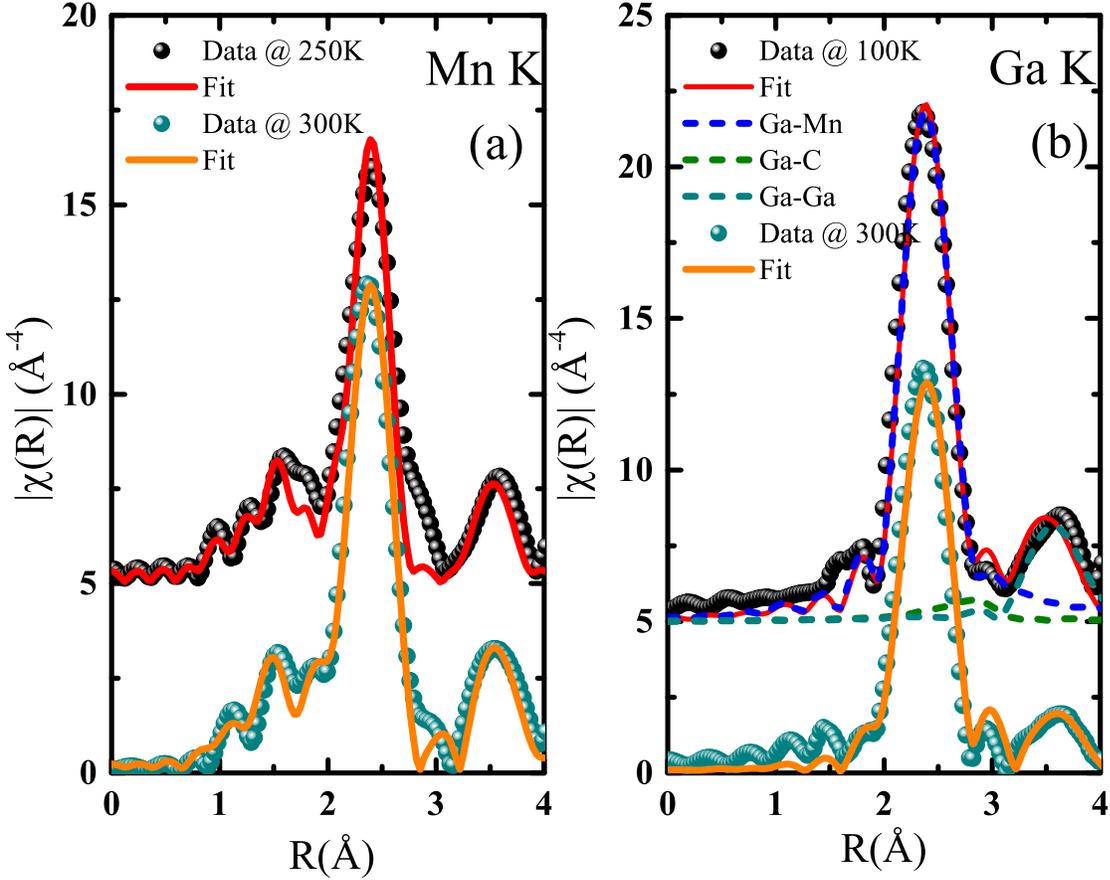}
\caption{k$^3$ weighted magnitude of Fourier transform of EXAFS ($|\chi (R)|$) along with best fitted lines at (a) Mn K edge and (b) Ga K edge at two different temperatures. Some of the data are scaled artificially along y-axis for clarity.}
\label{fig:exafs1}
\end{center}
\end{figure}

XAFS which is widely used as a tool to define the local structure about a specific atom was utilized to directly measure the temperature dependent local structure of Mn and Ga in Mn$ _{3} $GaC. Finer aspects of the technique were utilized to map structural distortions that may arise within the local structure of the corner sharing Mn$ _{6} $C octahedra at the first order transition. XAFS data were collected and analyzed up to $k$ = 16 \AA~$ ^{-1} $ ($\sim$ 1000 eV) from the absorption edge energy at the Mn and Ga K edges of Mn$ _{3} $GaC in the temperature range of 50 K to 300 K. In order to isolate contributions from the various near neighbor atoms, the oscillatory XAFS signals ($ \chi(k) $)  were appropriately weighted by $k^n$ ($n$ = 1, 2 or 3) and Fourier transformed (FT) to the $R$ - space  to yield $ \chi(R) $). The magnitude of $ \chi(R) $ ($|\chi(R)|$) of Mn and Ga K edge XAFS recorded at 300 K, depicted in Figure \ref{fig:exafs1}, presents peaks due to direct and multiple scattering from neighboring atoms around the absorbing atom. The position of these peaks are usually taken to give a measure of bond distance in case of direct scattering correlations. There is however a correction, due to phase difference introduced by both scattering and absorbing atom, that needs to be applied to get the correct bond distance or scattering path length. The position of the peaks in Figure \ref{fig:exafs1} are not corrected for any phase difference and hence appear at lower values than actual bond distances. The values of bond distances mentioned in supplementary text \cite{supp} are however phase corrected. The first peak in the $|\chi(R)|$ of Mn XAFS (Figure \ref{fig:exafs1}(a)) at $R \approx$ 1.5 \AA~ represents contributions from the first coordination shell containing the two carbon atoms (Mn-C) while the peak with highest intensity centered near 2.5 \AA~ contains contributions from equidistant Mn-Mn and Mn-Ga correlations. There is also a weak contribution from 90$^\circ$ Mn-C-Mn (referred to as Mn-C-Mn$\perp$) multiple scattering correlation to this peak. The next peak at about 3.5 \AA~ results from contributions due to direct scattering from third nearest neighbor, Mn (Mn-Mn(2)) as well as multiple scattering from linear Mn-C-Mn correlations which consist of Mn-C-Mn(2)-Mn and Mn-C-Mn(2)-C-Mn scattering paths. For clarity the geometry of the above mentioned multiple scattering paths is described in supplementary text \cite{supp}. Magnitude of FT of Ga K XAFS shows one strong peak due to contribution from the nearest 8 Mn atoms followed by a weak peak at about 2.8 \AA~ due to contribution from Ga-C scattering and another peak at 3.5 \AA~ due to Ga-Ga correlation. Both Mn and Ga XAFS data was fitted with all major structural correlations up to 4.0 \AA.

\begin{figure}
\begin{center}
\includegraphics[width=\columnwidth]{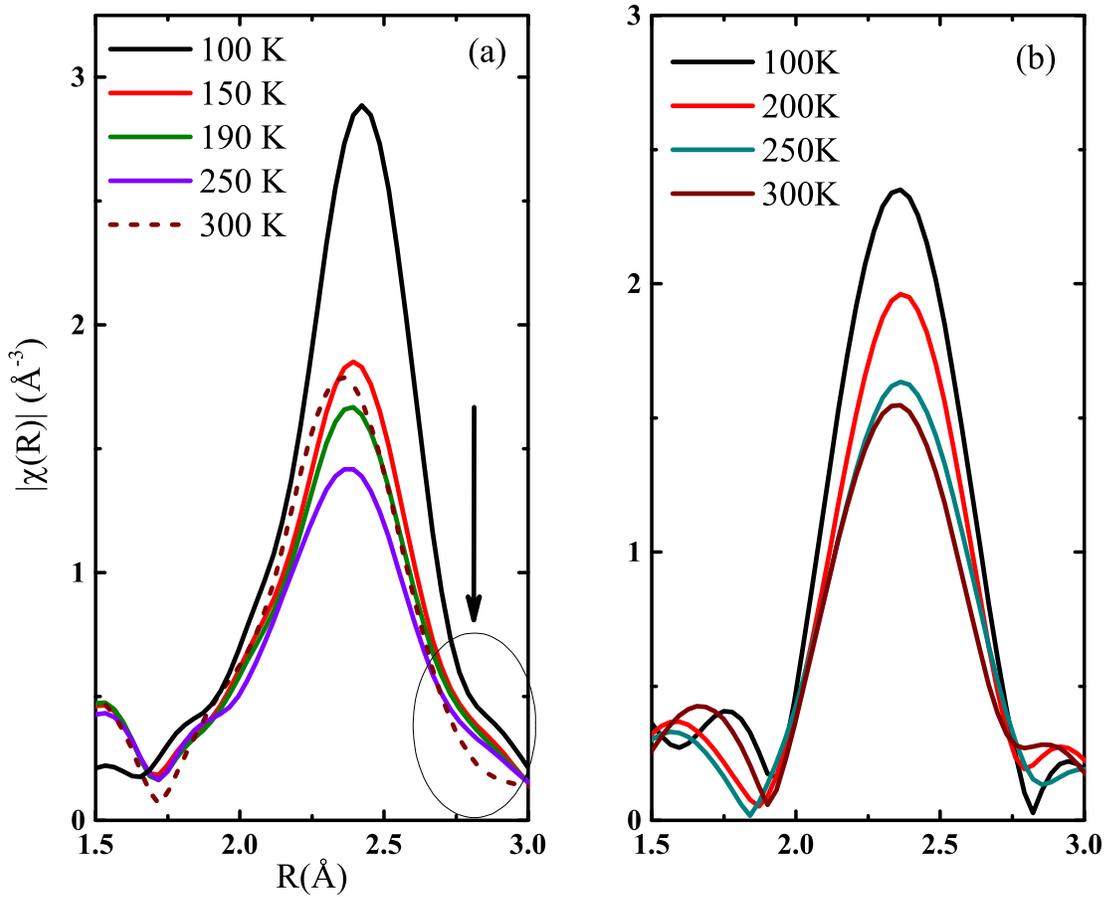}
\caption{Magnified image of magnitude of Fourier transform (FT) (|$\chi(R)|$) of (a) Mn K and (b) Ga K edge XAFS in the R range of 1.5 \AA~ to 3.0 \AA~ at a few representative temperatures. The circled area denoted by an arrow in (a) shows the growth of additional feature in FT of Mn XAFS below T = 300K.}
\label{fig:exafs2}
\end{center}
\end{figure}

In the first attempt, the local structures around Mn and Ga in Mn$_{3}$GaC were fitted using restrictions imposed by cubic symmetry (cubic model). A good fit was observed for Ga XAFS data at 300 K with bond distances in good agreement with values calculated from lattice parameter. In fact good fits were obtained at all temperatures down to 100 K with the cubic model as can be seen from Figure \ref{fig:exafs1}(b) which depicts $|\chi(R)|$ data of Ga XAFS at 300 K and 100 K. However, there were some issues with the best fit to Mn XAFS though it appeared to be good and gave quite a low R-factor ($\sim$ 0.02). The major issues with the fitting primarily were, (a) lower values of Mn-Mn and Mn-Ga bond distances as compared to their values from lattice parameter (see Table S1 of supplementary information) and (b) very high value (0.04 \AA$^2$) of mean square radial disorder ($\sigma^2$) for the Mn-C-Mn$\perp$ path. These discrepancies amplified further with decrease in temperature. A clear misfit between data and best fitted curve can be clearly seen just below 3 \AA~ for the 250 K data in Figure \ref{fig:exafs1}(a). Figure \ref{fig:exafs2}(a) presents a magnified image in the region of the second peak in $|\chi(R)|$ of Mn XAFS at some representative temperatures across the two magnetic transitions. It can be seen that distortions in the peak shape start appearing as the temperature is lowered to 250 K which is close to T$_C$ of Mn$_3$GaC and are present at all lower temperatures. No such discrepancies were noted in Ga XAFS as can be seen from the behavior of peak corresponding to Ga-Mn correlation plotted in Figure \ref{fig:exafs2}(b). It has been shown earlier in Figure \ref{fig:exafs1}(b) that the local structure around Ga confirmed with the lattice symmetry at all temperatures even below the phase transition temperature.

Based on the fact that local structure around Ga is in conformity with cubic symmetry and using the value of Ga-Mn distance in fitting Mn XAFS, the non conformity of Mn XAFS with cubic symmetry was identified to be due to distortions in the second near neighbour, Mn-Mn distance. In particular, the eight fold degenerate Mn-Mn correlation splits into two, Mn-Mn$_{long}$ and Mn-Mn$_{short}$, correlations with lengths of 3.1 \AA~ and 2.74 \AA~ respectively. In comparison, the Mn-Mn bond distance obtained from lattice parameter is 2.75 \AA. At 300 K the percentage of distortion is quite small (20$\pm$12\%). It increases slowly to about 50$\pm$5\% at 250 K and remains nearly the same at all lower temperatures. It may be noted that Mn$_3$GaC orders ferromagnetically at 242 K.

\begin{figure}
\begin{center}
\includegraphics[width=\columnwidth]{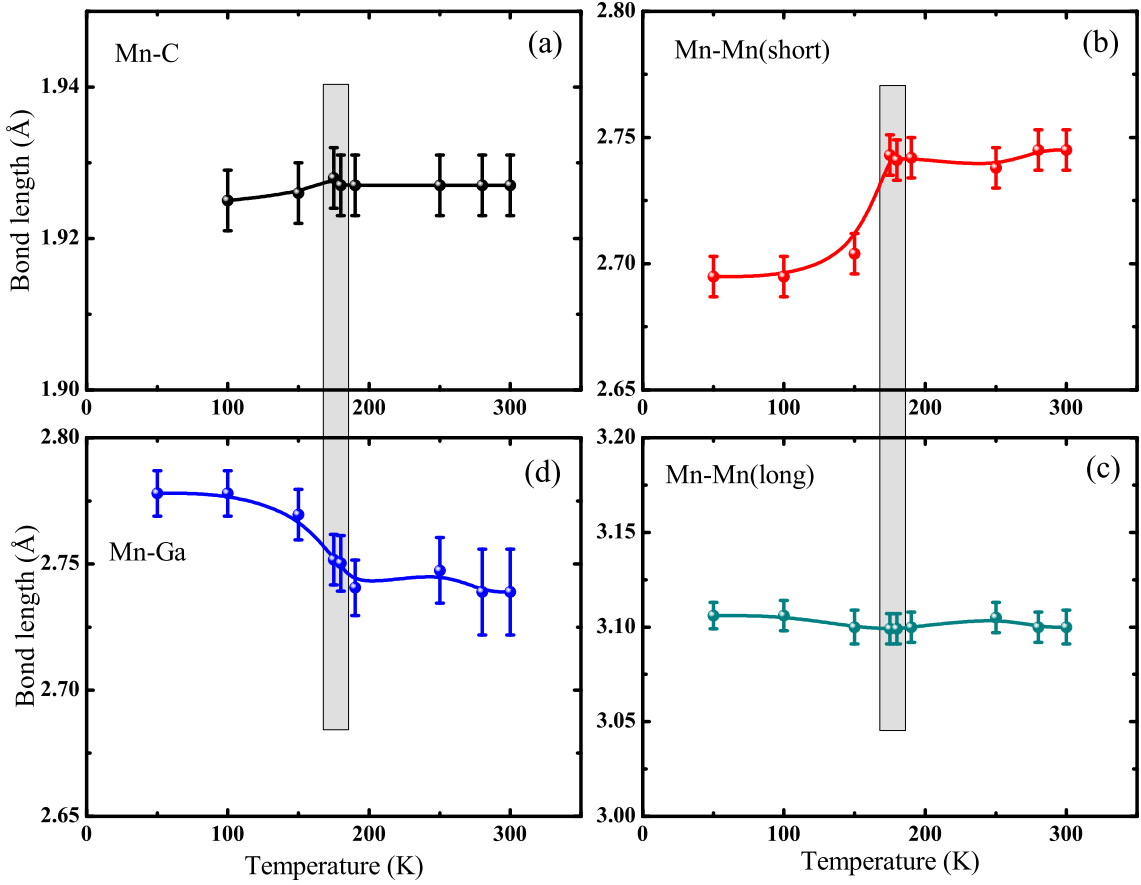}
\caption{Plot of variation of different bond distances obtained from analysis of Mn K and Ga K XAFS. (a) Mn-C bond distance, (b) Mn-Mn$_{short}$ bond distance, (c) Mn-Mn$_{long}$ bond distance and (d) Mn-Ga bond distance. While Mn-C and Mn-Mn bond distances have been obtained from Mn XAFS analysis, Mn-Ga bond distances have been obtained from Ga K XAFS analysis.}
\label{fig:exafs3}
\end{center}
\end{figure}

\begin{figure}
\begin{center}
\includegraphics[width=\columnwidth]{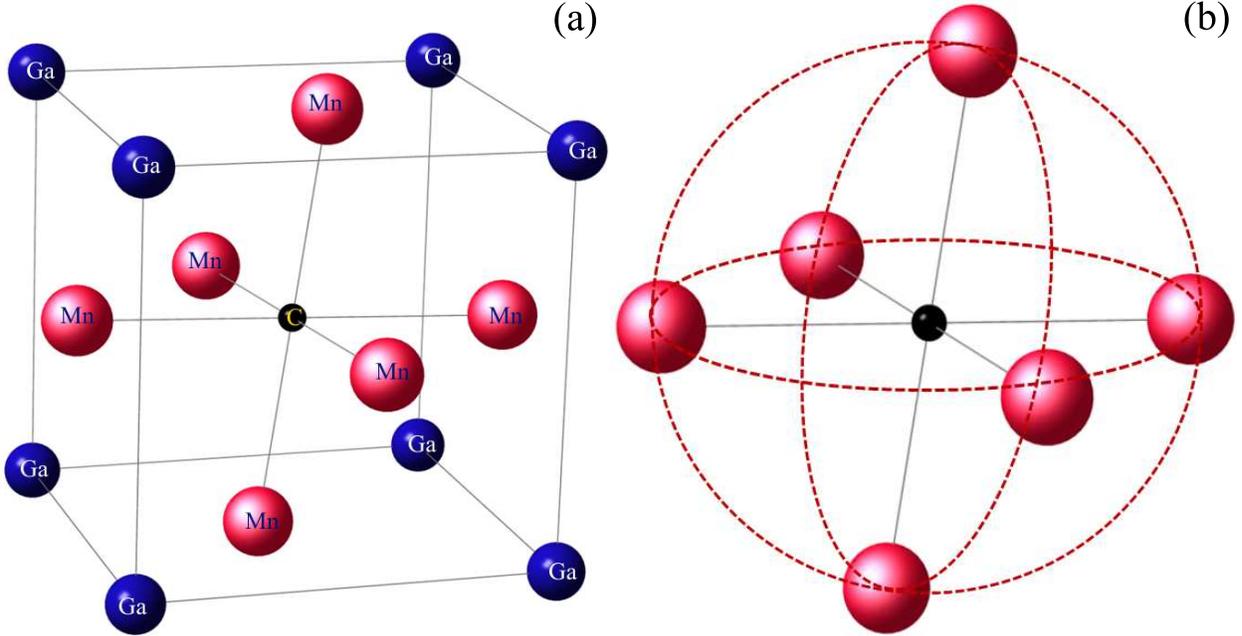}
\caption{A schematic representation of distortions of Mn sublattice within the unit cell of Mn$_3$GaC. (a) Mn$_3$GaC unit cell with Mn atoms displaced from their face centered positions. (b) Mn$_6$C octahedra with Mn atoms distorted along the circular arcs of radius equal to Mn-C bond distance.}
\label{fig:scheme}
\end{center}
\end{figure}

Another interesting observation that comes to fore from the analysis of Mn XAFS is the variation of Mn-C bond distance and the mean square disorder in its value. Despite the compound undergoing a discontinuous volume expanding phase transition, Mn-C bond distance remains nearly invariant in the entire temperature range. The variation of Mn-C bond length as a function of temperature is plotted in Figure \ref{fig:exafs3}(a). The values of $\sigma^2$ for this correlations are also less than 0.001\AA$^2$. Such small values of $\sigma^2$ indicate rigidity of the bond which is not entirely unexpected. Band structure calculations show that Mn $3d$ states are strongly hybridized with C $2p$ states near Fermi level \cite{Shim200266}. What is surprising however, is the lack of temperature dependence of both bond length and $\sigma^2$. Such a behavior of Mn-C bond distance indicates that it has very little or no role to play in the magneto-structural transition exhibited by Mn$_3$GaC. This is in agreement with C K XAS results reported earlier \cite{Lewis200618}. Furthermore, in light of the nearly invariant Mn-C bond distance, the distortions in the Mn-Mn bond distances imply movement of Mn atoms away from the face centered positions, but on a circular arc with a radius equal to Mn-C bond distance. Such a distortion will result in nearly equal number of long and short Mn-Mn bond distances. This agrees well with our observation of 50\% long and short Mn-Mn distances from XAFS analysis. A schematic depicting such distortions of the Mn sublattice is shown in Figure \ref{fig:scheme}.

The temperature variation of long and short Mn-Mn  bond distances are plotted in Figure \ref{fig:exafs3}(b) and (c) while Figure \ref{fig:exafs3}(d) depicts variation of Mn-Ga bond distance as a function of temperature. Mn-Ga bond distance shows very little variation in the temperature range 300 K to 190 K. Below 190 K there is an abrupt increase in its value coinciding with the volume expansion transition at 175 K followed again by a nearly constant value of bond distance below 150 K. The Mn-Mn bond distances on the other hand show completely different behavior. The Mn-Mn$_{short}$ bond length exhibits an abrupt decrease near the first order transition temperature while Mn-Mn$_{long}$ bond distance has nearly temperature independent variation in the entire range studied here. While the behavior of Mn-C, Mn-Ga and the two Mn-Mn bond distances are not in accordance with the variation of lattice parameter, Ga-Ga bond distance (which is actually equal to the lattice parameter $a$) obtained from XAFS analysis has a temperature variation very similar to that of $a$ obtained from x-ray diffraction, thus indicating that in Mn$_3$GaC, only the Mn sublattice is distorted while the overall cubic lattice symmetry is preserved by Ga and C atoms.

Goodenough Kanamori rules have been invoked to understand magnetism of Mn$_3$GaC. In particular, the the 90$^\circ$ Mn-Mn exchange interactions are believed to be responsible for ferromagnetism while the antiferromagnetic ordering is believed to occur due to 180$^\circ$ Mn-C-Mn exchange interactions.  However, the invariance of Mn-C bond distance across the phase transition temperature as seen from XAFS analysis raises serious questions on this hypothesis. Instead the splitting of Mn-Mn correlations in nearly equal number of long and short distances at around T$_C$ and the abrupt decrease in shorter Mn-Mn bond distance near the phase transition temperature can be used to explain the ferromagnetic and antiferromagnetic transitions exhibited by Mn$_3$GaC.

It is well known that Mn-Mn distances greater than 2.82 \AA~ support ferromagnetism while smaller distances ($\sim$ 2.72 \AA) aid in antiferromagnetic ordering \cite{RevModPhys.25.64}. Therefore ferromagnetic order in Mn$_3$GaC can occur due to Mn-Mn$_{long}$ distance. The appearance of structural distortions just above T$_C$ also lend support to this argument. Furthermore, the ferromagnetic order in Mn$_3$GaC is not stable, in fact the compound is reported to undergo intermediate magnetic transitions before transforming to antiferromagnetic state \cite{Kanomata199867}. The splitting of Mn-Mn bonds into long and short bonds at 3.1 \AA~ and 2.74 \AA~ could explain the presence of competing magnetic interactions. The Mn-Mn$_{short}$ distance abruptly decreases from 2.74 \AA~ to 2.69\AA~ near the first order magnetic transition temperature. This abrupt decrease results in strengthening of antiferromagnetic interactions and the compound orders antiferromagnetically.

The displacement of Mn atoms from their face centered position on a circular arc of radius equal to Mn-C bond distance leading to long and short Mn-Mn bonds is quite similar to rhombohedral distortion of the unit cell. However, in the present case the distortions are confined only to Mn sublattice. At T$_N$, these distortions are at their maximum and therefore to prevent a change in symmetry from cubic to rhombohedral, the lattice undergoes a cubic-cubic volume expansion transition. The increase in Mn-Ga bond distance below the volume expanding phase transition temperature indicates presence of strain in the Mn sub-lattice. This strain can be compensated using a compressive stress like hydrostatic pressure or magnetic field causing a reverse transformation to low volume ferromagnetic state.

\section*{Conclusion}
In conclusion, the Mn sublattice in Mn$_3$GaC undergoes a distortion with lowering of temperature such that the Mn atoms are displaced from their face centered positions on a circular arc of radius equal to Mn-C bond length. Such a distortion results in longer and shorter Mn-Mn bond distances. Magnetism of Mn$_3$GaC is controlled by RKKY type exchange interactions between the Mn-Mn$_{long}$ and Mn-Mn$_{short}$ bonds. While the longer Mn-Mn bond distances ($\sim$ 3.1 \AA~) are responsible for ferromagnetic ordering, the shorter Mn-Mn bond distances support antiferromagnetic interactions. An abrupt decrease in the value of Mn-Mn$_{short}$ distance from 2.74 \AA~ to 2.69 \AA~ at 175 K is responsible for antiferromagnetic ordering. In order to prevent a decrease in the lattice symmetry from cubic to a rhombohedral due to the abrupt decrease in Mn-Mn$_{short}$ distance, the compound undergoes a volume expanding cubic-cubic structural transition at the same temperature. The metamagnetic transition from AFM to FM under magnetic field or pressure then occurs due to alleviation of the strain caused by the decrease of 0.05 \AA~ in Mn-Mn$_{short}$ at the first order magnetic transition.

\section*{Acknowledgments}
This work is supported by Board of Research in Nuclear Sciences (BRNS) under the project 2011/37P/06. M/s Devendra D. Buddhikot, and Ganesh Jangam are acknowledged for their assistance in magnetization measurements. Authors thank Photon Factory, KEK, Japan for beamtime on beamlines 9C and NW10A for the proposal No. 2014G042. Parts of this research were carried out at the light source PETRA III at DESY, a member of the Helmholtz Association (HGF). We would like to thank Edmund Welter and Roman Chernikov for assistance in using beamline P-65. We would also like to thank the Department of Science and Technology, India for the financial assistance provided to conduct the experiments under the DST-DESY project.

\bibliographystyle{apsrev4-1}
\bibliography{Ga}

\end{document}